\documentclass[eqsecnum,preprint,preprintnumbers,12pt,aps]{revtex4}
\usepackage{epsfig,psfrag,subfigure}
\usepackage{color}

\begin{document}

\title{Renormalized  Effective Actions in Radially Symmetric
Backgrounds: Exact Calculations Versus Approximation Methods}
\author{Gerald V.~Dunne}
\email{dunne@phys.uconn.edu}
\affiliation{Department of Physics, University of Connecticut, Storrs, CT 06269, U.S.A.}
\author{Jin Hur}
\email{hurjin@kias.re.kr}
\affiliation{School of Computational Sciences, Korea Institute for
Advanced Study, Seoul 130-012, Korea}
\author{Choonkyu Lee}
\email{cklee@phya.snu.ac.kr}
\affiliation{Department of Physics and Astronomy and Center for Theoretical
Physics\\ Seoul National University, Seoul 151-742, Korea}
\author{Hyunsoo Min}
\email{hsmin@dirac.uos.ac.kr}
\affiliation{Department of Physics,  University of Seoul , Seoul 130-743, Korea}

\vspace{3cm}
\begin{abstract}
Our previously-developed calculational method   
(the partial wave cutoff method) is employed to
evaluate explicitly scalar one-loop effective actions in a class
of radially symmetric background gauge fields. Our method proves
to be particularly effective when it is used in conjunction with a
systematic WKB series for the large partial wave contribution to
the effective action. By comparing these numerically exact
calculations against the predictions based on the large mass
expansion and derivative expansion, we discuss the validity ranges
of the latter approximation methods.
\end{abstract}
\maketitle

\section{Introduction}\label{intro}
In field theoretic investigations one is often confronted by the
rather formidable task to evaluate the one-loop effective action
in some nontrivial background field. Until recently, it has not
been possible in four spacetime dimensions to evaluate explicitly this
renormalized quantity (including its full finite part), unless some
very special background is chosen or {\it a priori} arbitrary
parameters (e.g., mass values) are set to zero. In our recent
publications \cite{idet, radial} we made some headway to this old
problem by developing an efficient calculational method -- a
combination of analytic and numerical schemes -- for the exact
computation of fully renormalized one-loop effective actions in radially symmetric
backgrounds.   For example, this method was first  applied to the accurate
determination of QCD single-instanton determinants for arbitrary
quark mass values \cite{idet}, producing a result that interpolates smoothly between 
the known analytical massless and heavy quark limits.   In Ref.\cite{radial} 
we generalized the calculational procedure to calculate the one-loop 
effective action in {\it any} radially
symmetric background, not just an instanton.
In the present
paper, which is a sequel to Ref.\cite{radial}, we present
some   explicit examples and 
results (including the numerical contributions) to establish 
the efficiency and generality of our method.
We also examine the validity of often-used approximation methods,
such as the large mass expansion and the derivative expansion, compared to
numerically exact calculations.

In Ref.\cite{radial} we derived some relevant formulas needed in
the calculation of the scalar one-loop effective action (in
Euclidean spacetime), assuming SU(2) background gauge fields of the
form
\begin{eqnarray}
\label{acase1} &\text{(Case 1): }& A_{\mu}({\bf x}) = 2
\eta_{\mu\nu a} x_\nu
f(r) \frac{\tau^a}{2}, \\
\label{acase2} &\text{(Case 2): }& A_{\mu}(\mathbf{x}) = 2
(\eta_{\mu\nu i} \hat{u}_i) x_\nu g(r) \frac{\tau^3}{2},
\end{eqnarray}
where $\mu,\nu = 1,2,3,4$, $r \equiv \sqrt{|\mathbf{x}|} =
\sqrt{x_\mu x_\mu}$, $\eta_{\mu\nu a}$ (or $\eta_{\mu\nu i}$) are the 
't Hooft symbols \cite{thooft}, and $\hat{u}_i$ a unit 3-vector.
Case 1 is inherently non-Abelian, while Case 2 has a fixed color direction and so is quasi-Abelian. These backgrounds are characterized by the radial profile functions $f(r)$ and $g(r)$, respectively. In each case the spectral problem separates into partial waves due to the spherical symmetry.

Our method has been deliberately developed so that it can accommodate {\it numerical} input for $f(r)$ and $g(r)$, since this situation often arises in quantum field theory applications. But,  to illustrate the method more clearly, 
 we choose here specific {\it Ans\"atze} for the radial profile functions.
In Case 1, of   `non-Abelian type' (\ref{acase1}), we choose the radial function $f(r)$ of the form
\begin{equation}\label{backf}
f(r)= \frac{1}{r^2} H(r)\quad , \quad H(r)= \frac{(r/\rho)^{2\alpha}}{1+(r/\rho)^{2\alpha}}\quad ,
\end{equation}
with free parameters $\rho$ and $\alpha$ (under the  regularity
restriction     $|\alpha|\geq 1$).   [The BPST instanton solution
\cite{belavin} corresponds to the choice $\alpha=1$]. 
In Case 2, of the quasi-Abelian type (\ref{acase2}), we choose the radial function $g(r)$ of the form
\begin{equation}\label{backg}
g(r)= B \left\{ 1-\tanh \left[ \beta (\sqrt{B}\,r-\xi_0) \right]
\right\}\quad ,
\end{equation}
with three free parameters $\beta$, $\xi_0$ and $B$ (all taken to be
positive). In the limit $\beta \rightarrow 0$, (\ref{acase2})
then approaches the case of uniform field strength $B$. 
For finite $\beta$, (\ref{acase2})  represents a spherical bubble type potential
with radius $\frac{\xi_0}{\sqrt{B}}$ and wall thickness $\sim \frac{1}{\beta\sqrt{B}}$.

In this paper we calculate the renormalized scalar
one-loop effective action (including its full finite part) in the gauge field background pertaining
to the above two types.   The effective action is computed for arbitrary choices of the 
parameters characterizing the shape of the background field, not relying on the background field being slowly or rapidly varying, or on the particle mass being large or small relative to the scales set by the background field.
In performing this analysis, we have
found   that   significantly greater calculational efficiency   and precision  can be attained by
making a systematic use of   higher-order quantum mechanical WKB-type
approximations \cite{wkbpaper} for  the large 
partial-wave contributions to the effective action. This permits the extension of the high partial wave radial-WKB approximation to lower and lower partial waves, and results in dramatic numerical improvements.  Finally, we compare our results to the
predictions based on the large mass expansion and the derivative expansion.
To our knowledge, this kind of genuinely unambiguous
comparison in four-dimensional gauge theory has not been made
before.

This paper is organized as follows. In Sec. \ref{section2} we give
a short outline of our numerically exact calculational scheme and
also collect, for later use in the paper, relevant formulas from the
large mass expansion and derivative expansion for the scalar
one-loop effective action. Our detailed study on the one-loop
effective action with non-Abelian-type backgrounds (\ref{acase1})
is then presented in Sec. \ref{section3}. This is followed in Sec.
\ref{section4} by the corresponding study with quasi-Abelian
backgrounds (\ref{acase2}); in this case, the one-loop effective action is essentially that of scalar QED. In Sec. \ref{section5} we conclude
with some related discussions and comments.

\section{Brief Summary of our Calculational Scheme and Other
Approximation Methods}\label{section2}

The calculational method developed in Refs.
\cite{idet,radial} can be summarized and streamlined as follows.   For scalar fields in
radially symmetric non-Abelian background gauge fields $A_\mu({\bf
x})$, it is possible to express the corresponding (Euclidean)
one-loop effective action as a sum of individual partial-wave
contributions, i.e., $\sum_{J=0}^\infty \Gamma_J(A;m)$, with the
$J$-partial-wave term  [including the appropriate degeneracy factor]  given by radial functional determinants
\begin{equation}
\Gamma_J(A;m) = \ln \left( \frac{\det(-\tilde{D}_J^2
+m^2)}{\det(-\tilde{\partial}_J^2 +m^2)} \right)
\end{equation}
or, equivalently, by the proper time representation
\begin{equation}
\Gamma_J(A;m) = - \int_0^\infty \frac{ds}{s} e^{-m^2s}
\int_0^\infty dr \; \text{tr} \left\{ \tilde{\Delta}_J(r,r;s) -
\tilde{\Delta}^{\text{free}}_J(r,r;s) \right\}.
\end{equation}
Here, $m$ is the scalar mass, $-\tilde{D}_J^2$
($-\tilde{\partial}_J^2$) denotes the quadratic {\it radial}
differential operator relevant to the $J$-partial-wave in the
given background (or with $A_\mu(\mathbf{x})$ set to zero), and
$\tilde{\Delta}_J(r,r;s)$, $\tilde{\Delta}^{\text{free}}_J(r,r;s)$
represent the coincidence limits of the related proper-time Green
functions specified by the radial `heat' equations:
\begin{eqnarray}
&& \left[ \partial_s - \tilde{D}_{J \; (r)}^2 \right]
\tilde{\Delta}_J(r,r';s) = 0, \qquad (s>0) \\
&& s \rightarrow 0+: \qquad \tilde{\Delta}_J(r,r';s)
\longrightarrow \delta(r-r'). \nonumber
\end{eqnarray}
(The tildes over differential operators and Green's  functions imply
that we are here considering reduced operators/functions after
taking out various radial measure factors \cite{radial}). While the
quantities $\Gamma_J$ are finite individually, the partial wave
series $\sum_{J=0}^\infty \Gamma_J$ diverges and should be
renormalized. We thus introduce the intermediate partial wave
cutoff $J_L$ (a large, but finite value of $J$) and express the
fully renormalized effective action by two separate terms
\begin{equation}\label{Jsum}
\Gamma_{\text{ren}}(A;m) = \Gamma_{J\leq J_L}(A;m) +
\Gamma_{J>J_L}(A;m),
\end{equation}
where
\begin{eqnarray}
\Gamma_{J\leq J_L}(A;m) &\equiv& \sum_{J=0}^{J_L} \Gamma_J(A;m) = \sum_{J=0}^{J_L}
\ln \left( \frac{\det(-\tilde{D}_J^2
+m^2)}{\det(-\tilde{\partial}_J^2 +m^2)} \right), \label{finitepart} \\
\Gamma_{J>J_L}(A;m) &=& \lim_{\Lambda \rightarrow \infty} \left[ -
\int_0^\infty \frac{ds}{s} \left( e^{-m^2s} - e^{-\Lambda^2s}
\right) \sum_{J>J_L} F_J(s)
 \right. \nonumber \\
 &&  \left. - \frac{1}{12} \frac{1}{(4\pi)^2} \ln \left(
\frac{\Lambda^2}{\mu^2} \right) \int d^4\mathbf{x} \; \text{tr} \left(
F_{\mu\nu} F_{\mu\nu} \right) \right]\label{infinitepart}
\end{eqnarray}
with
\begin{equation}\label{FJ}
F_J(s)=\int_0^\infty dr\, \text{tr} \left( \tilde{\Delta}_J(r,r;s)
- \tilde{\Delta}^{\text{free}}_J(r,r;s) \right).
\end{equation}
Notice that the renormalization counter-term is incorporated in the second term $\Gamma_{J>J_L}(A;m)$; $\mu$ is the normalization mass, and $F_{\mu\nu}\equiv i[D_\mu, D_\nu]$ (with $D_\mu \equiv \partial_\mu -iA_\mu$) denotes the background Yang-Mills field strengths. The first term, $\Gamma_{J\leq J_L}$, may be evaluated numerically using the Gel'fand Yaglom method \cite{gy,kleinert,gvd} for
one-dimensional functional determinants. As for the second term
$\Gamma_{J>J_L}$, analytically developed expressions valid for
large enough $J_L$ were used in Refs. \cite{idet,radial}; for
this, we made essential use of the quantum mechanical radial WKB
expansion and the Euler-Maclaurin summation formula.   A crucial observation is that even though 
the partial-wave sum in (\ref{finitepart}) is quadratically
divergent as we let $J_L \rightarrow \infty$, this divergence
is {\it exactly cancelled} by a similar term originating from the
second term $\Gamma_{J>J_L}$ in (\ref{infinitepart}) in the large $J_L$-limit.   Thus, for
the sum of the two terms, we secure a finite result in the large
$J_L$-limit -- this is the essence of our numerically accurate scheme
for the determination of the renormalized effective action.   The ``low'' partial wave contribution is computed numerically, and the ``high'' partial wave contribution is computed analytically in the large $J_L$ limit using radial WKB. 
The ultraviolet cutoff dependence appears in the {\it analytic} high partial wave computation, and this allows standard renormalization techniques to be used, leading to the finite renormalized effective action.
For details on our renormalization prescription (and on how to go
from ours to other prescriptions), see Refs. \cite{idet,radial,kwon}.

Clearly, numerical efficiency of our calculational scheme hinges on how quickly the $J_L\to\infty$ limit is attained;  that is, 
on how much we can lower our partial wave cutoff $J_L$ to secure a reliable large-$J_L$ limit value for the above sum of the two terms. 
 (Note that if we were able to calculate both parts, i.e., $\Gamma_{J\leq J_L}$ and $\Gamma_{J>J_L}$ {\it exactly}, their sum would be independent of the choice of the cutoff value $J_L$.  An interesting example of this situation is the effective action for {\it massless} quarks in a single BPST instanton background, where our technique reproduces exactly and analytically \cite{idet} 't Hooft's result \cite{thooft} 
$\tilde{\Gamma}(m=0)\equiv\alpha\left(\frac{1}{2}\right)=-\frac{17}{72}-\frac{1}{6}\ln\,2+\frac{1}{6}-2\zeta^\prime(-1) \approx 0.145873...$.). In \cite{idet}, we chose the cutoff value $J_L$ to be rather large 
(of the order of 50), and used the Richardson extrapolation method \cite{bender} to reduce the numerical round-off error. There, it was sufficient to use the expression for $\Gamma_{J>J_L}$ with $O(\frac{1}{J_L})$ or smaller terms suppressed.   But, in the present,  more extensive, study, we have found that the computation is greatly improved, both in computing time and precision, by including [analytically calculated] higher-order WKB terms in the expression for $\Gamma_{J>J_L}$, so that it becomes valid up to the $O(\frac{1}{J_L^4})$ accuracy. These higher-order WKB terms can be found straightforwardly from   the $\frac{1}{l}$-expansion \cite{radial,lee2} for the radial proper-time Green function. With this procedure, we were able to ensure that the renormalized effective action we calculate is independent of the (moderately large) $J_L$-value with relative error of order $10^{-6}$, which is comparable to the final total error involved in our numerical computation of (\ref{finitepart}). 
In the end, we obtain comparable precision with much lower values of $J_L$, of the order of 10 or 20. This improved precision is demonstrated in sections \ref{section3} and \ref{section4} for  the backgrounds in (\ref{acase1}) and (\ref{acase2}).

We also give the results of some popular approximation schemes for the one-loop effective action, which are supposed to capture accurate values in certain limits. First, we have the large mass expansion \cite{novikov,kwon} of the scalar one-loop effective action which is obtained most easily with the help of the Schwinger-DeWitt proper-time expansion (or heat kernel expansion). Here, from the renormalized effective action, it is convenient to separate $\mu$-dependent pieces (and use dimensional considerations) to write
\begin{eqnarray}
\label{renscale1} &\text{(Case 1): }& \Gamma_{\rm ren}(A;m)= \frac{1}{6} \ln \left(\mu\rho\right)\int \frac{d^4 \mathbf{x}}{(4\pi)^2} \text{tr}\, F_{\mu\nu}^2
+ \tilde\Gamma(m\rho) \\
\label{renscale2} &\text{(Case 2): }& \Gamma_{\rm ren}(A;m)= \frac{1}{6} \ln \left(\frac{\mu}{\sqrt{B}}\right)\int \frac{d^4 \mathbf{x}}{(4\pi)^2}\text{tr}\, F_{\mu\nu}^2
+ \tilde\Gamma \left(\frac{m}{\sqrt{B}} \right)
\end{eqnarray}
where dimensionless constants are not indicated in an explicit manner. The quantity $\tilde{\Gamma}(m\rho)$ or $\tilde{\Gamma}(\frac{m}{\sqrt{B}})$, which is independent of $\mu$, can be expressed for large enough $m$ by an asymptotic series of the following form
\begin{eqnarray}
\label{largemass1} && \tilde{\Gamma}(m\rho)= -\frac{1}{6}\ln(m\rho)\int \frac{d^4 \mathbf{x}}{(4\pi)^2}\text{tr}F_{\mu\nu}^2 + \sum_{n=3}^\infty \frac{(n-3)!}{(m^2)^{n-2}} \int\frac{d^4\mathbf{x}}{(4 \pi)^2}\text{tr}\, a_n (\mathbf{x}, \mathbf{x}),\\
\label{largemass2} && \tilde\Gamma \left( \frac{m}{\sqrt{B}} \right)=  -\frac{1}{6}\ln \left( \frac{m}{\sqrt{B}} \right) \int \frac{d^4 \mathbf{x}}{(4\pi)^2}\text{tr}F_{\mu\nu}^2
+ \sum_{n=3}^\infty \frac{(n-3)!}{(m^2)^{n-2}} \int\frac{d^4\mathbf{x}}{(4 \pi)^2}
\text{tr}\, a_n (\mathbf{x}, \mathbf{x}).
\end{eqnarray}
Here $a_n(\mathbf{x},\mathbf{x})$, $n=3,4,5,\cdots$ denote appropriate coefficient functions in the Schwinger-DeWitt expansion: explicitly, for the traces of the $a_3$- and $a_4$-terms, we have \cite{novikov,kwon}
\begin{eqnarray}\label{a3}
\text{tr}\,a_3 (\mathbf{x}, \mathbf{x})&=& -{1\over 6} \ \text{tr} \left[
 i \frac{2}{15}  F_{\kappa\lambda}F_{\lambda\mu}F_{\mu\kappa}
-\frac{1}{20} (D_\kappa F_{\lambda\mu}) (D_\kappa F_{\lambda\mu})
\right],
\\
\text{tr}\, a_4 (\mathbf{x}, \mathbf{x})
 &=&{1\over 24} \ \text{tr}\left[- \frac{1}{21} F_{\kappa\lambda}F_{\lambda\mu}
F_{\mu\nu}F_{\nu\kappa}
+  \frac{11}{420} F_{\kappa\lambda}F_{\mu\nu}F_{\lambda\kappa} F_{\nu\mu}
 +  \frac{2}{35} F_{\kappa\lambda}F_{\lambda\kappa}F_{\mu\nu}F_{\nu\mu}
\right.\nonumber \\
&~&+  \frac{4}{35} F_{\kappa\lambda}F_{\lambda\mu}F_{\kappa\nu}F_{\nu\mu}
 + i \frac{6}{35}  F_{\kappa\lambda}(D_\mu F_{\lambda\nu})(D_\mu F_{\nu\kappa})
 + i \frac{8}{105}  F_{\kappa\lambda}(D_\lambda F_{\mu\nu})
(D_\kappa F_{\nu\mu})
\nonumber \\
 &~&\left. +  \frac{1}{70} (D_\kappa D_\lambda F_{\mu\nu})(D_\lambda D_\kappa F_{\nu\mu})
\right], \label{a4}
\end{eqnarray}
where $D_\lambda F_{\mu\nu} \equiv [D_\lambda, F_{\mu\nu}]$ and $D_\kappa D_\lambda F_{\mu\nu} \equiv [D_\kappa,[D_\lambda, F_{\mu\nu}]]$, etc.    Note that in these expressions for $a_3(\mathbf{x}, \mathbf{x})$ and $a_4(\mathbf{x}, \mathbf{x})$, we have {\it not} assumed that $A_\mu(\mathbf{x})$ satisfies the classical equations of motion. We also comment that while this large mass expansion is rather simple to use, it is in fact an asymptotic expansion, and so its regime of useful applicability is restricted to the mass being large relative to $\frac{1}{\rho}$ or $\sqrt{B}$, respectively.   See Sec. \ref{section3} and Sec. \ref{section4} for comparisons of the large mass expansions (\ref{largemass1}) and (\ref{largemass2}) with our exact numerical answers.

There is another well-known approximation method to the one-loop effective action, the derivative expansion. The leading order of the derivative expansion corresponds to using the Euler-Heisenberg constant field result \cite{heisenberg,duff,dunne-review}, but substituting the inhomogeneous fields for the homogeneous ones used to compute the Euler-Heisenberg effective action. This approximation is very simple to implement, and  is expected  to be a good approximation when the spacetime variation in the background gauge field strengths is sufficiently `slow' so that we may regard their derivatives as small terms in the effective action. Subleading derivative expansion contributions can also be computed, but we will not consider them here. A systematic study  of the validity range of this method is still lacking  (although the Borel summability properties of the derivative expansion have been analyzed in a nontrivial soluble inhomogeneous QED example in \cite{cangemi}). 
We remark here that  if the background gauge fields are genuinely
non-Abelian (as in our Case 1), the convergence character of the related expansion --
the so-called covariant derivative expansion \cite{leutwyler,yildiz,gargett,salcedo} -- is
less certain. In this paper we restrict ourselves to
applying the derivative expansion with our quasi-Abelian backgrounds
(\ref{acase2}) only. In that case, the
leading term in the derivative expansion is given by the Euler-Heisenberg formula \cite{heisenberg,duff,dunne-review}
\begin{eqnarray}\label{dexpansion}
\tilde\Gamma_{\rm DE}(\frac{m}{\sqrt{B}})=&-& \int dr\frac{r^3}{4} \int_{0}^\infty \frac{d s}{s^3}
e^{-m^2 s}\left(\frac{E_1 s}{\sinh(E_1 s)}\frac{E_2 s}{\sinh(E_2 s)}-1 +\frac{s^2}{6} (E_1^2+E_2^2)\right)  \nonumber \\
&-&\frac{1}{6}\ln \left( \frac{m}{\sqrt{B}} \right) \int \frac{d^4 \mathbf{x}}{(4\pi)^2}\text{tr}F_{\mu\nu}^2,
\end{eqnarray}
where $\pm  i E_1$ and $\pm i E_2$ denote four eigenvalues of the $4\times 4$ matrix $\mathbf{F}=(F_{\mu\nu})$, with
\begin{equation}
E_1 =\frac{1}{2} \sqrt{{\cal F}- \sqrt{{\cal F}^2- {\cal G}^2}}, \qquad E_2 =\frac{1}{2} \sqrt{{\cal F}+\sqrt{{\cal F}^2-{\cal G}^2}},
\end{equation}
where
\begin{equation}
{\cal F} = \frac{1}{2} \text{tr}F_{\mu\nu}F_{\mu\nu}, \qquad {\cal G} = \frac{1}{4}\text{tr} \epsilon_{\mu\nu\lambda\kappa}F_{\mu\nu}F_{\lambda\kappa}.
\end{equation}
Consult Sec. \ref{section4} to see how the predictions based on this method fare against the accurate numerical calculations.

\section{Non-Abelian backgrounds}\label{section3}
\subsection{Properties of the Background Fields}
\begin{figure}
\psfrag{hor}{$\frac{r}{\rho}$}
\psfrag{ver}{$H(r)$}
%\subfigure[Plot]{\includegraphics[width=0.7\linewidth]{solFplus.eps}}\\
\subfigure[]{\includegraphics[width=0.7\linewidth] {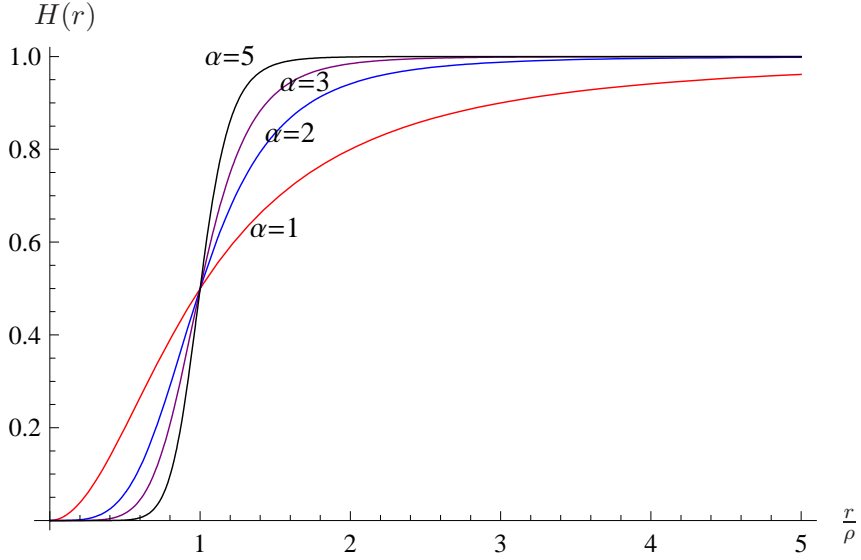}}
\psfrag{hor}{$\frac{r}{\rho}$}
\psfrag{ver}{$H(r)$}
\subfigure[]{\includegraphics[width=0.7\linewidth] {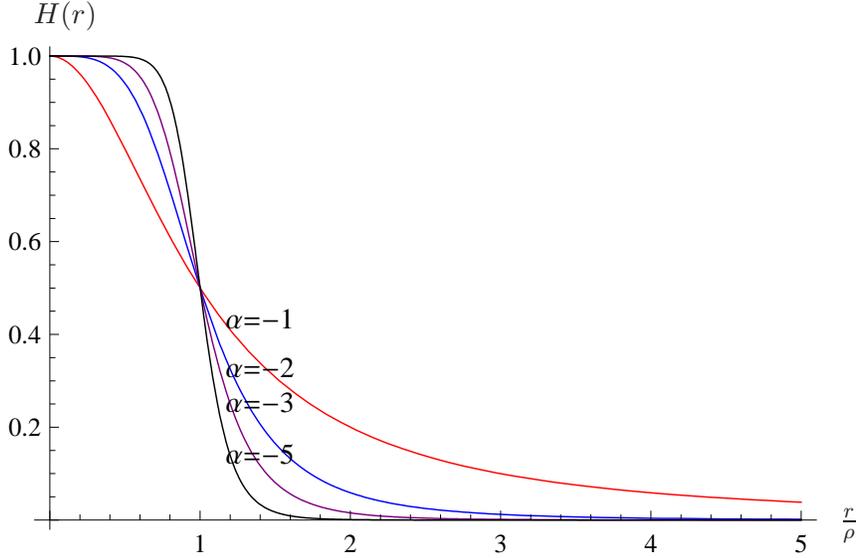}}
\caption{Plots of the radial profile function $H(r)$ appearing in the non-Abelian gauge field (\ref{acase1}), as a function of $r/\rho$. We have drawn the cases with $\alpha=1,2,3,5$ in (a), and the cases with $\alpha=-1, -2, -3, -5$ in (b). Note that $H(r)$ behaves like a step function if $|\alpha|$ becomes very large, with the step localized at $r=\rho$.}
\label{fig1}
\end{figure}

We consider here a family of radial background fields described by (\ref{acase1}) and (\ref{backf}), which resemble the single instanton configuration. The parameter $\alpha$ is chosen to be in the range $|\alpha| \geq 1$, so that $A_\mu (\mathbf{x})$ is well-behaved at the origin $r=0$. When $\alpha$ takes a negative value it is convenient to cast the function $H(r)$ in the form
\begin{equation}\label{hfunctionneg}
H(r)=\frac{1}{1+(r/\rho)^{2|\alpha|}}.
\end{equation}
Note that while our configuration with $\alpha=1$ is simply the single instanton solution in the regular gauge \cite{idet}, by choosing $\alpha=-1$ the single anti-instanton solution in the {\it singular} gauge is also obtained. In Fig. \ref{fig1}, we have plotted the shape of the background function $H(r)$ for several values of  $\alpha$'s.

For the gauge field (\ref{acase1}), the corresponding  field strength tensor $F_{\mu\nu}$ is
\begin{eqnarray}\label{fmunu}
F_{\mu\nu}=\frac{\tau^a}{2} \left( 4\eta_{\mu\nu a}\frac{H(H-1)}{r^2}+2(x_\mu\eta_{\nu\lambda a}-x_\nu\eta_{\mu\lambda a})\frac{x_\lambda}{r^4}\left[2H(H-1)+r H^\prime\right] \right)\quad ,
\end{eqnarray}
where $H^\prime\equiv \frac{d}{dr}H(r)$.   Alternatively, inserting the expression (\ref{backf}) for $H(r)$, we find
\begin{equation}\label{fmunu-h}
F_{\mu\nu}=\frac{\tau^a}{2} \left( -\eta_{\mu\nu a}-(x_\mu\eta_{\nu\lambda a}-x_\nu\eta_{\mu\lambda a})\frac{x_\lambda}{r^2}(1-\alpha ) \right)\frac{4(r\rho)^{2\alpha}}{r^2(\rho^{2\alpha}+r^{2\alpha})^2} \quad .
\end{equation}
Note that the field strength tensor $F_{\mu\nu}$ has a singularity at $r=0$ when $|\alpha|<1$, and the second part of (\ref{fmunu-h}), the part proportional to $(x_\mu\eta_{\nu\lambda a}-x_\nu\eta_{\mu\lambda a})x_\lambda$,  vanishes if $\alpha=1$ (i.e. for the single instanton solution). For these instanton-like radial background fields the corresponding classical action is readily evaluated:
\begin{eqnarray}
\frac{1}{2} \!\int d^4\!\mathbf{x}\, \text{tr}\, F_{\mu\nu}^2&=& 12\pi^2 \int_{0}^{\infty} \frac{d r}{r}[r^2 (H^\prime)^2+ 4 H^2(H-1)^2] \nonumber \\
&=& 4\pi^2 \frac{1+\alpha^2}{|\alpha|}
\end{eqnarray}
Note that this result does not depend on the scale parameter $\rho$, and has a minimum value when $\alpha=\pm 1$, i.e., for the single instanton or single anti-instanton solution.

For the general vector potential considered here, the topological winding number is
\begin{eqnarray}
\frac{1}{32\pi^2}\epsilon_{\mu\nu\lambda\tau}  \int d^4\!\mathbf{x}\, \text{tr}\,  F_{\mu\nu} F_{\lambda\tau} =- \int_0^\infty \! dr \; \frac{d}{d r}( 2 H^3-3 H^2 ) = \pm 1.
\end{eqnarray}
This means that all the configurations corresponding to positive values of $\alpha$ belong to the class of winding number $1$, and those corresponding to  negative values of $\alpha$ to the class of winding number $-1$. 
Further, notice that self-dual configurations occur when $\frac{1}{2} \!\int d^4\!\mathbf{x}\, \text{tr}\, F_{\mu\nu}^2=\frac{1}{2} \int d^4\!\mathbf{x}\, \text{tr}\,  F_{\mu\nu}\tilde{F}_{\mu\nu}$, which means $\alpha^2=1$, corresponding to the BPST instanton or anti-instanton for which $\alpha=\pm 1$.

\subsection{Large Mass Expansion}
Before delving into the  exact calculation of the one-loop effective action,  we first present the result of the large mass expansion. In our background fields we find for the traces of leading coefficient functions $a_3(\bf{x},\bf{x})$ and $a_4(\bf{x},\bf{x})$ (see (\ref{a3}) and (\ref{a4})), the following explicit results:
\begin{eqnarray}
\int d^4 \mathbf{x} \; \text{tr}\,a_3 &=& \int_0^\infty \frac{dr}{\rho^2}\frac{r^{6\alpha-3}}{15  \left(1+r^{2 \alpha }\right)^6} \left[
\left(10-13 \alpha ^2-33 \alpha ^4\right)    \right. \nonumber \\
&&\left.+r^{2 \alpha } \left(5+20 \alpha +22 \alpha ^2+28 \alpha ^3+21 \alpha ^4\right)+ (\alpha\to -\alpha)
\right], \\
\int d^4 \mathbf{x}\; \text{tr}\,a_4 &=& \int_0^\infty \frac{dr}{\rho^4} \frac{r^{8 \alpha -5}}{210  \left(1+r^{2 \alpha }\right)^8} \left[
4 r^{4 \alpha } (1+\alpha )^2 \left(28+112 \alpha +105 \alpha ^2+188 \alpha ^3+107 \alpha ^4\right)  \right. \nonumber\\
&&  -8 r^{2 \alpha } \left(-91-273 \alpha -21 \alpha ^2+309 \alpha ^3+618 \alpha ^4+1164 \alpha ^5+694 \alpha ^6\right)\nonumber\\
&& \left. +\frac{1}{2} \left(1337-5250 \alpha ^2-3399 \alpha ^4+11992 \alpha ^6\right) +(\alpha \to -\alpha) \right].
\end{eqnarray}
Then, after performing the $r$-integration, the large-mass expansion for the one-loop effective action is seen to take the form (we here give the result for the quantity $\tilde{\Gamma}(m\rho)$, introduced in (\ref{renscale1}))
\begin{eqnarray}
\tilde{\Gamma}_{\rm LM}(m\rho)&=&-\frac{(1+\alpha^2)}{12 |\alpha|} \ln(m\rho)
+\frac{\pi(5-10\alpha^2+11\alpha^4-6\alpha^8)}{1800 \alpha^6\sin(\pi/\alpha)}\frac{1}{(m\rho)^2}
+(\alpha^2 -4) (\alpha^2 -1 ) \nonumber \\
&&\times \frac{\pi\left(-140+35 \alpha ^2-378 \alpha ^4-317 \alpha ^6+120 \alpha ^8\right)}{88200 \alpha^8\sin(2\pi/\alpha)} \frac{1}{(m\rho)^4}+\cdots. 
\label{gammalm}
\end{eqnarray}
Note that, for $|\alpha|=1$ (or $2$), taking the limit $|\alpha|\to 1$ (or $|\alpha|\to 2$) in the right hand side of (\ref{gammalm}) should be understood.
This large mass expansion result (\ref{gammalm}) will be compared with the numerically determined effective action later.

We make a comment on the length scale parameter $\rho$ here. The modified effective action $\tilde{\Gamma}(m\rho)$ does not depend on the renormalization mass scale $\mu$, and so is a function only of the dimensionless combination $m\rho$. We may then set the size parameter $\rho=1$ during the calculation, without loss of generality, and readily restore it in the final result.

\subsection{Numerically Accurate Calculation of the Lower Angular Momentum Part}
We now turn to our accurate effective action calculation based on (\ref{Jsum}). First consider the lower angular momentum part $\Gamma_{J\leq J_L}$, given in (\ref{finitepart}). Note that, in the present backgrounds, the partial waves are specified by the quantum numbers $J=(l, j, j_3,\bar{l}_3)$, as described in detail in Ref. \cite{radial}. We are working with isospin $\frac{1}{2}$, so $j=l\pm \frac{1}{2}$. Using the notation of \cite{radial}, the radial Hamiltonian, representing $-D^2$ in the given partial wave sector, assumes the form
\begin{equation}
{\cal H}_{l,j} \equiv -D^2_{(l,j)}=-\frac{\partial^2}{\partial r^2}-\frac{3}{r} \frac{\partial}{\partial r} +V_{l,j}
\end{equation}
with
\begin{equation}
V_{l,j}(r)=\frac{4l(l+1)}{r^2}+ \left[j(j+1)-l(l+1)-\frac{3}{4}\right] \frac{4 H(r)}{r^2} +3 \frac{H^2(r)}{r^2},
\end{equation}
while in the absence of the background field
\begin{equation}
{\cal H}^{\rm free}_{l} \equiv -\partial^2_{l}=-\frac{\partial^2}{\partial r^2}-\frac{3}{r} \frac{\partial}{\partial r} + \frac{4l(l+1)}{r^2}\quad .
\end{equation}
The radial Hamiltonian is independent of the quantum numbers $j_3$ and $\bar{l}_3$; this introduces the degeneracy factor $(2j+1)(2l+1)$ in the partial wave sum below.

Having identified the relevant quantum numbers, the lower angular momentum part $\Gamma_{J\leq J_L}$ can be written as (here, $L$ serves as our partial wave cutoff)
\begin{equation}\label{low}
\Gamma_{J\leq J_L} (A;m) =\sum_{l=0}^{L}(2l+1)\sum_{j=l-1/2}^{l+1/2}(2j+1)  \ln  \left( \frac{{\det(\cal H}_{l,j}
+m^2)}{\det({\cal H}_l^{\rm free} +m^2)} \right)\quad ,
\end{equation}
The ratio of two determinants in (\ref{low}) is determined, according to the Gel'fand-Yaglom method \cite{gy,kleinert,gvd}, by the ratio of the asymptotic
values of two wave functions as
\begin{eqnarray}
\frac{\det({\mathcal H}_{l,j}+m^2)}{\det({\mathcal H}^{\mathrm { free}}_l+m^2)}=\lim_{R\to\infty}\left(\frac{\psi_{l,j}(R)}{\psi^{\rm free}_l(R)}\right).
\label{theorem}
\end{eqnarray}
Here $\psi_{l,j}(r)$ and $\psi^{\rm free}_l(r)$ denote the solutions to the radial differential equations
\begin{eqnarray}
({\mathcal H}_{l,j}+m^2) \, \psi_{l,j}(r)&=&0, \label{ode} \\
({\mathcal H}^{\rm free}_{l}+m^2) \, \psi^{\rm free}_l(r)&=&0, \label{freeode}
\end{eqnarray}
which have the same small-$r$ behaviors, i.e.,
\begin{equation}
r\to 0: \;\; \psi_{l,j}(r)\sim r^l, \; \qquad \psi^{\rm free}_{l}(r)\sim r^l.
\end{equation}

Note that  the solution to (\ref{freeode}), which is the modified Bessel
function
\begin{eqnarray}
\psi^{\rm free}_{l}(r)=\frac{I_{2l+1}(m r)}{r}\quad ,
\label{besseli}
\end{eqnarray}
grows exponentially fast at large $r$, as do the
numerical solutions to (\ref{ode}) for the operators ${\mathcal
H}_{l,j}+m^2$. Thus, numerically, it is advantageous to consider the  {\it ratio}, $\psi_{l,j}(r)/\psi^{\rm free}_l(r)$, which stays finite for all $r$. In fact, since we compute the logarithm of the determinant, we can directly consider the logarithm of the ratio:
\begin{eqnarray}
{S}_{(l,j)}(r)=\ln \frac{\psi_{(l,j)}(r)}{\psi^{\rm free}_{(l)}(r)},
\label{ratio}
\end{eqnarray}
which also has a finite value in the large $r$ limit.
This function satisfies the differential equation
\begin{eqnarray}
&& \frac{d^2 S_{(l,j)}}{dr^2}+\left(\frac{d S_{(l,j)}}{dr}\right)^2+\left(\frac{1}{r}+2m\frac{I^\prime_{2l+1}(m r)}{I_{2l+1}(m r)}\right)\frac{d S_{(l,j)}}{dr}=U_{(l,j)}(r)\quad ,\label{nonlinear} \\
&& U_{(l,j)}(r)=V_{l,j} -\frac{4l(l+1)}{r^2} \nonumber \\
&&\phantom{U_{(l,j)}(r)}=\frac{4j(j+1)-4l(l+1)-3}{r^2(1+r^{2\alpha})}+\frac{3}{r^2(1+r^{2\alpha})^2},
\label{potential}
\end{eqnarray}
under the initial value boundary conditions
\begin{eqnarray}
S_{(l,j)}(r=0)=0\qquad , \qquad S^\prime_{(l,j)}(r=0)=0\quad .
\label{logbc}
\end{eqnarray}
Noting that the eigenvalues of the total angular momentum $j$ equal $l\pm \frac{1}{2}$ for a given value of $l$, it is convenient to combine the contributions $S_{(l,l+\frac{1}{2})}(r)$ and $S_{(l+\frac{1}{2},l)}(r)$, which come with the same degeneracy factor $(2l+1)(2l+2)$. With this understanding, it is possible to express the amplitude (\ref{low}) in the form
\begin{eqnarray}
\Gamma_{J\le J_L}(A;m)&=&\sum_{l=0,\frac{1}{2},1, \dots}^L (2l+1)(2l+2)P(l)\quad ,
\label{lj1}\\
P(l)&\equiv& S_{(l,l+\frac{1}{2})}(\infty)+S_{(l+\frac{1}{2},l)}(\infty)\quad .  \label{pwave}
\end{eqnarray}
Here $S_{(l,l+\frac{1}{2})}(\infty)$ and $S_{(l+\frac{1}{2},l)}(\infty)$ denote the asymptotic (i.e., $r\to\infty$) limits to the solutions of the differential equations in (\ref{nonlinear}) with the potentials
\begin{eqnarray}
V_{l,l+1/2}(r)=\frac{4l(l+1)}{r^2}+(4l+3) \frac{H(r)}{r^2} + 3\frac{H(r)(H(r)-1)}{r^2}\quad ,
\label{potential+}
\end{eqnarray}
and
\begin{eqnarray}
V_{l+1/2,l}(r)=\frac{4l(l+1)}{r^2}+(4l+3) \frac{(1-H(r))}{r^2} + 3\frac{(H(r)-1)H(r)}{r^2}\quad ,
\label{potential-}
\end{eqnarray}
respectively.

Note that the potentials $V_{l,l+1/2}(r)$ and $V_{l+1/2,l}(r)$ have the same form except that $H(r)$ in one expression gets replaced by $(1-H(r))$ in the other. For $H(r)$ given in (\ref{backf}), we have $(1-H(r))=\frac{1}{(1+r^{-2\alpha})}$,  while $H(r)=\frac{1}{(1+r^{2 \alpha})}$; i.e., the same expression as $(1-H(r))$ only with $\alpha$ in the latter replaced by $-\alpha$. Therefore, if we consider the effective action with the background parameter $\alpha$ replaced by $-\alpha$, the only change is that two potentials $V_{l, l+1/2}(r)$ and $V_{l+1/2,l}(r)$ are interchanged, and so the two quantities $S_{(l,l+\frac{1}{2})}(\infty)$ and $S_{(l+\frac{1}{2},l)}(\infty)$ in (\ref{pwave}) are also interchanged. This shows that each partial wave contribution to the effective action with the background parameter $-\alpha$ is the same as the one with the parameter $\alpha$ and thus two effective actions with $\alpha$ and with $-\alpha$ have the same value. (This was true for the!
 classical action also). Similar behaviors, concerning the cases with $\alpha=\pm 1$, were observed already in Ref. \cite{idet}. Based on this observation, consideration of the effective action for positive values of $\alpha$ is sufficient.

\begin{figure}
\psfrag{hor}{$r$}
\psfrag{ver}{$S_{(l,l+1/2)}(r)$}
\subfigure[Plot of the solutions $S_{l,l+1/2}(r)$ for $\alpha=1,2,3,4,5$ (from the bottom)]{\includegraphics[width=0.7\linewidth]{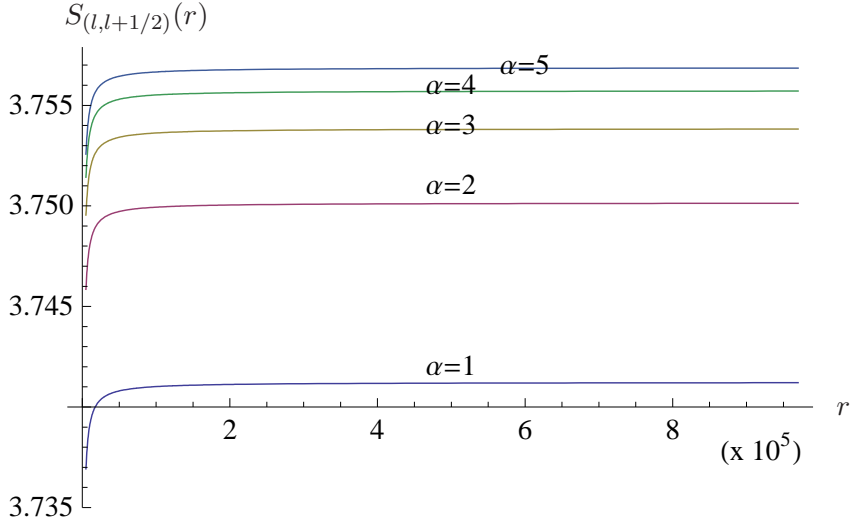}}\\
\psfrag{hor}{$r$}
\psfrag{ver}{$S_{(l+1/2,l)}(r)$}
\subfigure[Plot of the solutions $S_{l+1/2,l}(r)$ for $\alpha=1,2,3,4,5$ (from the top)]{\includegraphics[width=0.7\linewidth]{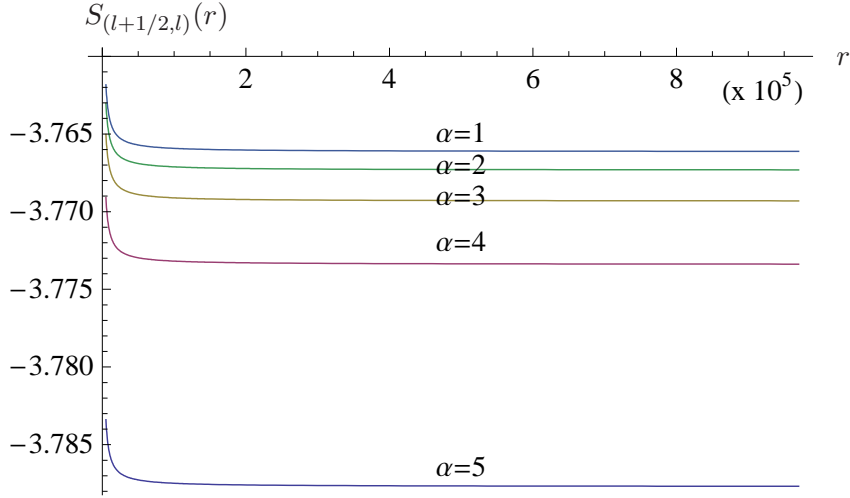}}
\caption{Plots of $S_{l,l+1/2}(r)$ and $S_{l+1/2,l}(r)$ when $m=1$ and $l=10$. Note that the asymptotic values of $S_{l+1/2,l}(r)$ and $S_{l,l+1/2}(r)$ are roughly of the same magnitude but with opposite signs.}
\label{solfig1}
\end{figure}

\begin{figure}
\psfrag{hor}{$l$}
\psfrag{ver}{$P(l)$}
\includegraphics[width=0.8\linewidth]{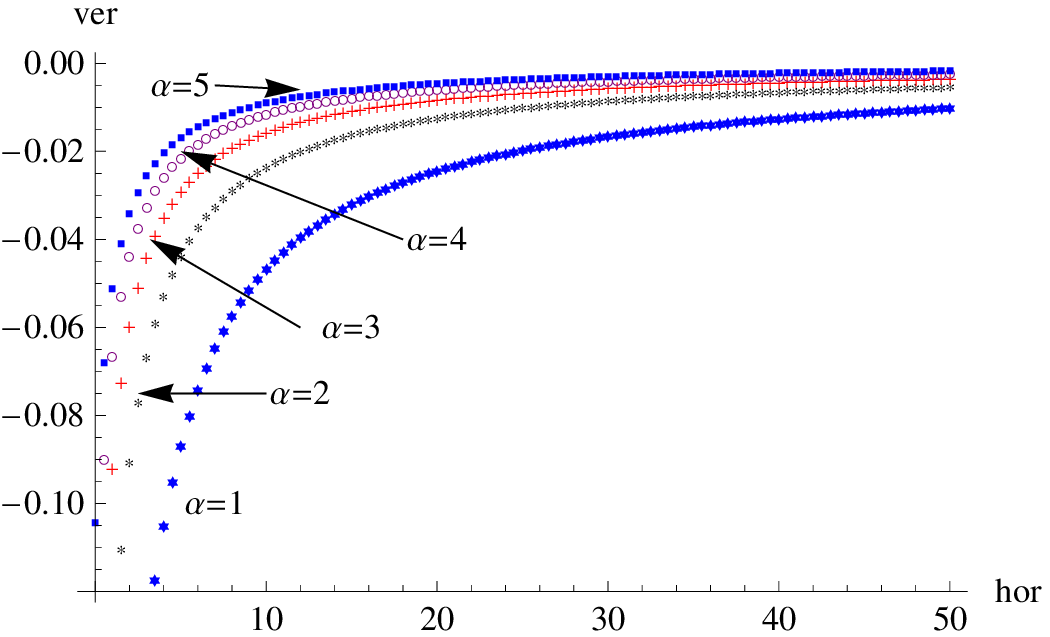}
\caption{Plots of partial wave contributions as a function of $l$ for the cases with $\alpha=1,2,3,4,5$ (from the bottom) and $m=1$.  }\label{partialfig}
\end{figure}

The ODE system specified by (\ref{nonlinear})-(\ref{logbc}) can easily be solved numerically. (With $m=0$ and $\alpha=\pm 1$, analytic solution to this equation was found in \cite{idet}). In Fig. \ref{solfig1} we plot the solutions for a few cases. It clearly shows that the solutions approach constant values in the $r\to\infty$ limit.
In Fig. \ref{partialfig} we plot  partial wave contributions with $l=0,1/2,\cdots$ for $m=1$ and $\alpha=2$.
Note that $P(l)\sim O(\frac{1}{l})$ when the angular momentum $l$ becomes large.
Since the degeneracy factor $(2l+1)(2l+2)$ is quadratic, this implies that $\Gamma_{J\leq J_L}$ in (\ref{lj1}) behaves as $L^2$ in the large $L$ limit.
This divergent behavior will be canceled when we add the higher angular momentum contribution.

\subsection{WKB Calculation of the Higher Angular Momentum Part}
We now calculate the higher angular momentum part $\Gamma_{J>J_L}$, given in (\ref{infinitepart}). This cannot be computed numerically (as we have done for $\Gamma_{J\le J_L}$) because very large partial wave
contributions lead naively to a divergent result and require careful renormalization to ensure a finite result. The large partial wave contribution depends on the regulating cutoff $\Lambda$, whose effect must be identified and isolated for renormalization; and this cannot easily be done numerically.
However, this quantity (incorporating renormalization) can be calculated {\it analytically} in a WKB-type asymptotic series, assuming that the partial wave cutoff $J_L$ is large enough. Here the higher angular momentum sum of the partial wave heat kernel, $\sum_{J>J_L}F_J(s)$ with $F_J(s)$ given by (\ref{FJ}), may be described more explicitly by the form 
\begin{eqnarray}
\sum_{J=J_L+\frac{1}{2}}F_{J}(s)&=& \int_0^\infty dr
\sum_{l=L+\frac{1}{2}}^\infty
(2l+1)(2l+2)
 \left\{ \tilde{\Delta}_{(l,l+\frac{1}{2})} (r,r;s) +
\tilde{\Delta}_{(l+\frac{1}{2},j=l)} (r,r;s) \right. \nonumber \\
&& -\left.  {\tilde\Delta}_{(l)}^{\rm free} (r,r;s) - {\tilde
\Delta}_{(l+\frac{1}{2})}^{\rm free} (r,r;s) \right\}.
\label{heatkexp}
\end{eqnarray}
Now, as explained in \cite{radial}, we may use the $\frac{1}{l}$ expansion for the modified radial proper-time Green function when $l$ is large.
When $\tilde{\Delta}(r,r;s)$ is expanded in terms with increasing number of derivatives of the potential, the scaling is such that when the $l$ sum is approximated by the Euler-Maclaurin formula, this generates the large $L$ expansion.
For a generic radial potential $V(r)$, this expansion has the following form: 
\begin{eqnarray} \label{green-largel}
\tilde{\Delta}(r,r;s) &=& \frac{1}{\sqrt{4\pi s}} e^{-s V(r)} \left\{
1 + \left( \frac{1}{12} s^3 (V')^2 - \frac{1}{6} s^2 V'' \right) \right. \nonumber \\
&+& \left( \frac{1}{288} (V')^4 s^6-\frac{11}{360} (V')^2 V'' s^5+\frac{1}{40} (V'')^2 s^4+\frac{1}{30} V' V^{(3)} s^4-\frac{1}{60} V^{(4)} s^3 \right)\nonumber \\
&+&\left(\frac{(V')^6 s^9}{10368}-\frac{17 (V')^4 V'' s^8}{8640}+\frac{83 (V'V'')^2 s^7}{10080}+\frac{1}{252} (V')^3 V^{(3)} s^7-\frac{61}{15120} (V'')^3 s^6 \right.   \nonumber \\
&& - \frac{43}{2520}  V' V'' V^{(3)} s^6
-\frac{5}{1008}(V')^2 V^{(4)} s^6+\frac{23}{5040} (V^{(3)})^2 s^5+\frac{19}{2520} V'' V^{(4)} s^5  \nonumber \\
&&\left.\left. +\frac{1}{280} V' V^{(5)} s^5-\frac{1}{840} V^{(6)} s^4\right)
+ O\left(\frac{1}{l^8}\right) \right\}.
\end{eqnarray}
Note that the terms are collected according to the total number of derivatives on $V$.
[In addition to the terms already calculated in (3.16) of \cite{radial}, we have here included some higher order terms as well because they will be useful in finding the large $J_L$ expansion of $\Gamma_{J>J_L}$]. The large-$L$ series expression for $\Gamma_{J>J_L}$ can then be found by inserting (\ref{green-largel}) into (\ref{heatkexp}) with the generic potential $V$ replaced by the potential $V_{l,j}$ or $V^{\rm free}_{l}$ in (\ref{potential+}) or (\ref{potential-}).
The summation over $l$ in (\ref{heatkexp}) can be performed using the Euler-Maclaurin summation method. The result is tantamount to the systematic WKB series, as was shown in Ref. \cite{radial}. Since this procedure was described already in Appendix C of \cite{radial}, we will not repeat it here. The final result in the present potential can be presented as a $\frac{1}{L}$ series of the form
\begin{eqnarray}
\Gamma_{J>J_L} &=& \int_0^\infty dr \left\{
Q_2(r) L^2 + Q_{1}(r) L + Q_{\rm log}(r) \ln L + Q_{0}(r) + Q_{-1}(r)
\frac{1}{L} +\cdots \right\}, \label{largeL}
\end{eqnarray}
where
\begin{eqnarray}
&& Q_{2}(r) = \frac{8 H(H-1)}{r\sqrt{\tilde{r}^2+4}}, \\
&& Q_{1}(r) = \frac{8(3\tilde{r}^2+8)}{r(\tilde{r}^2+4)^{3/2}} H(H-1), \\
&& Q_{{\rm log}}(r) = -\frac{1}{4r}\left(4H^2(H-1)^2 + r^2 H'^2\right), \\
&& Q_{0}(r) = \frac{1}{6r(\tilde{r}^2+4)^{7/2}} \left[
4(3\tilde{r}^6+49\tilde{r}^4+236\tilde{r}^2+352) H^2(H-1)^2 \nonumber
\right. \\
&& \qquad - 6(22\tilde{r}^6+157\tilde{r}^4+352\tilde{r}^2+384) H(H-1) +
16r(\tilde{r}^4+5\tilde{r}^2+4)(2H-1)H' \nonumber \\
&& \qquad \left. +r^2(\tilde{r}^2+4)^2 \left\{ (3\tilde{r}^2+8)H'^2 - 4(2H-
1)H'' \right\} \right] - Q_{{\rm log}}(r) \ln \left( \frac{\mu
r}{\sqrt{\tilde{r}^2+4} +2} \right), \qquad \\
&& Q_{-1}(r) = \frac{1}{r(\tilde{r}^2+4)^{9/2}}
\left[ 2(9\tilde{r}^6+36\tilde{r}^4-64\tilde{r}^2-256) H^2(H-1)^2 \right. \nonumber \\
&& \qquad +(-6\tilde{r}^8+25\tilde{r}^6+368\tilde{r}^4+128\tilde{r}^2) H(H-1) +8r\tilde{r}^2(\tilde{r}^4+3\tilde{r}^2-4)(2H-1)H' \nonumber \\
&& \qquad \left. -2r^2(\tilde{r}^2+4)^2 \left\{ 2(\tilde{r}^2+2)H'^2 + \tilde{r}^2(2H-1)H'' \right\} \right],
\end{eqnarray}
with $\tilde{r}= \frac{m r}{L}$. The $r$-integration, with $H(r)$ given in (\ref{backf}), can be performed numerically. 
The next higher order terms, which are quite lengthy and so are not given here, 
can also be calculated in a straightforward manner. In fact we have calculated the quantity $\Gamma_{J>J_L}$ up to $O(\frac{1}{L^4})$-terms, 
and the details of this calculation will be reported elsewhere \cite{hur}.

\subsection{Results for the Total One-loop Effective Action}
\begin{figure}
\begin{tabular}{cc}
\subfigure[\label{partsum1}]{\epsfig{file=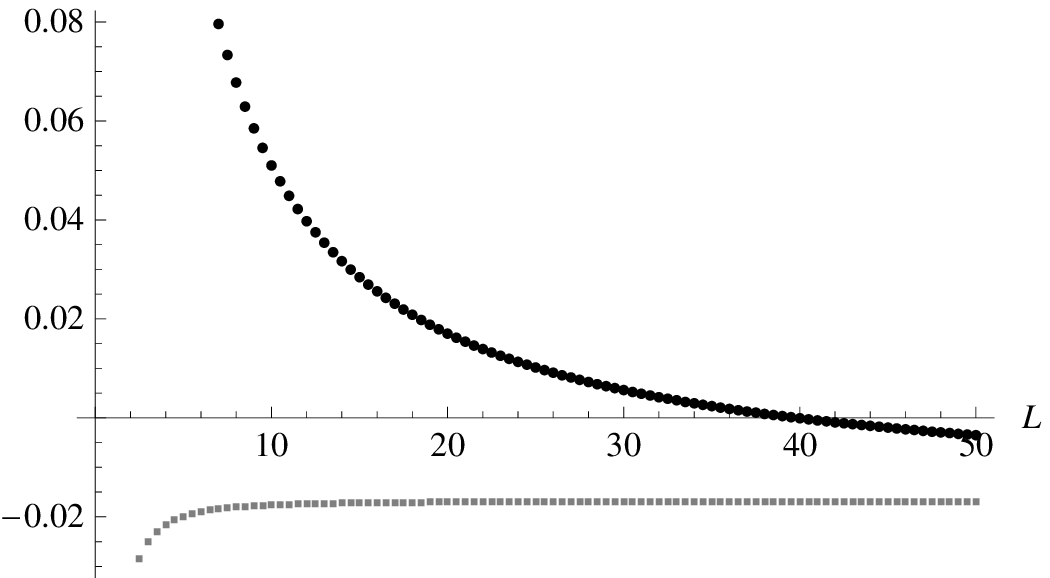,width=0.7\linewidth}} \\
\subfigure[\label{partsum2}]{\epsfig{file=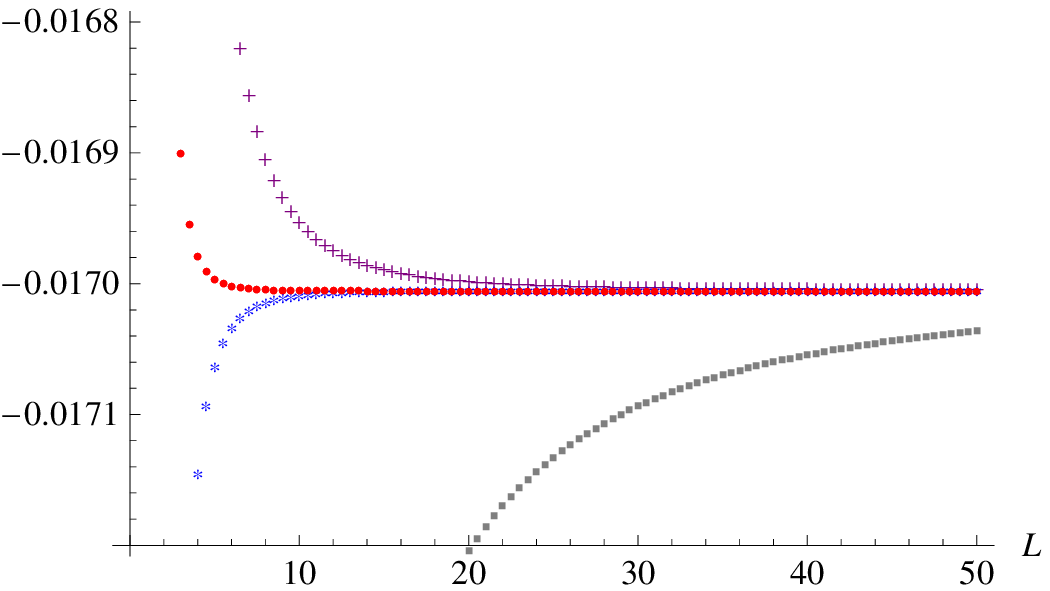,width=0.7\linewidth}}
\end{tabular}
\caption{We have plotted $L$-dependence of the sum of lower and higher angular momentum parts.  In (a) the upper (black) dots denote the case with all 
$O\left(\frac{1}{L}\right)$ terms ignored in $\Gamma_{J>J_L}$, and slow convergence is evident. the lower (grey) squares the results after incorporating $O\left(\frac{1}{L}\right)$ corrections, show better convergence.
In (b) the (grey) squares, (purple) crosses, (blue) stars and (red) dots represent the cases obtained after we incorporate $8\frac{1}{L}$, $\frac{1}{L^2}$, $\frac{1}{L^3}$ and $\frac{1}{L^4}$ corrections successively.}
\end{figure}

Let us now put together the lower and higher angular momentum parts, (\ref{lj1}) and (\ref{largeL}) respectively, of  the effective action, computed separately in the previous two subsections. Even if from the expression in (\ref{largeL}) we keep only up to the terms of $O(L^0)$ (i.e., up to the $Q_{0}$-term in the integrand) and add it to the lower angular momentum part found numerically in the subsection C, we obtain a finite result in the limit of very large $L$, with divergent contributions from the two parts canceling each other. However, for moderately large $L$ the resulting sum shows dependence on the cutoff value $L$. In other words, although the desired effective action should result if $L$ is taken to be sufficiently large, the rate of convergence is quite slow (see Fig.\ref{partsum1}).
This causes a problem in numerical efficiency.

One can accelerate the convergence by using the Richardson extrapolation method as described, say, in Ref.\cite{bender}, and this was in fact the method we used in the calculation of the instanton determinant in \cite{idet}. In this work we adopt a different, theoretically far more satisfying, approach to this problem. As we emphasized in Sec.\ref{section2}, the effective action should not depend on the choice of our cutoff value $L$ if the exact results for both $\Gamma_{J\leq J_L}$ and $\Gamma_{J>J_L}$ were used. This implies that, leaving aside possible numerical inaccuracy in calculating $\Gamma_{J\leq J_L}$, the $L$-dependence in the sum for finite cutoff value $L$ is really due to our ignoring $\frac{1}{L}$-suppressed contributions in the WKB series
 (\ref{largeL}). So we can systematically improve the large $L$ limit by adding higher order terms in the $\frac{1}{L}$-series for the higher angular momentum part $\Gamma_{J>J_L}$. As we mentioned already, we have identified the $\frac{1}{L}$-series up to terms of $O(\frac{1}{L^4})$. When these higher-order terms of the $\frac{1}{L}$-series are utilized, the situation changes dramatically: this is exhibited in Fig.\ref{partsum2}. From the figure we see that it is possible to achieve an $L$-independent result even for relatively lower and lower values of $L$ if we include higher and higher order terms in $\frac{1}{L}$ for the large angular momentum part. 
Even $L\sim 10$ produces good convergence. This reduces the number of numerical computations to be done in the low partial wave piece (\ref{lj1}). Thus, use of the systematic WKB series for $\Gamma_{J>J_L}$ plays a pivotal role in reducing the computer time in our calculation.

\begin{figure}
\psfrag{hor}{$m\rho$}
\psfrag{ver}{$\tilde\Gamma(m\rho)$}
\includegraphics[width=0.9\linewidth]{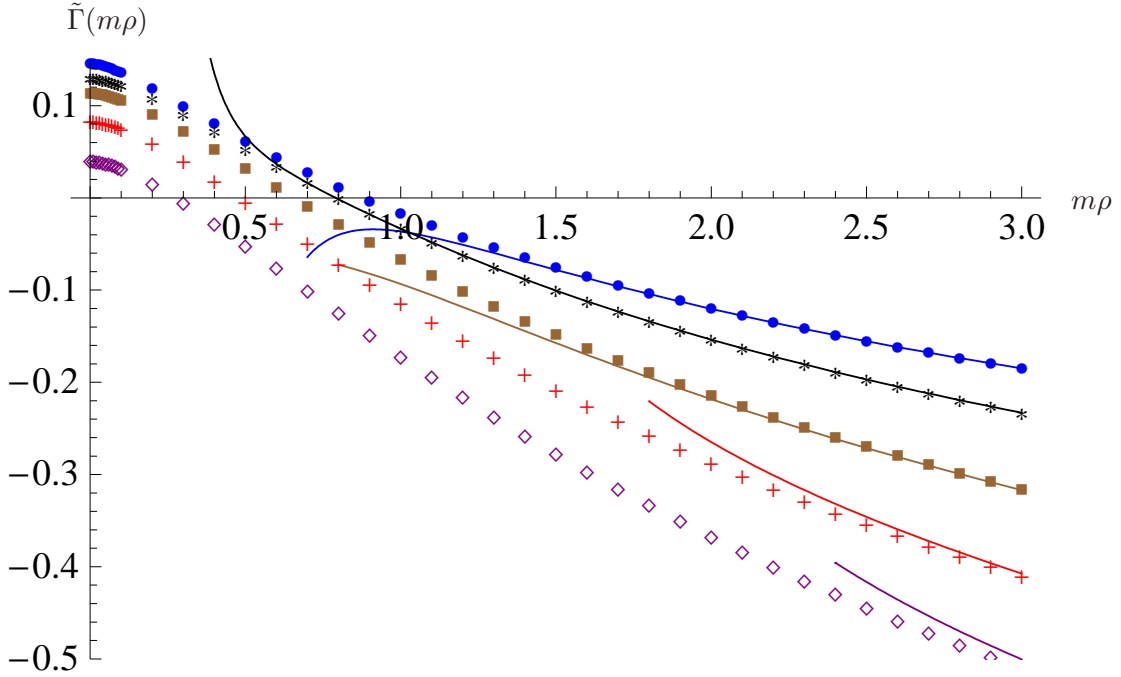}
\caption{Plots of the modified effective action as a function of $m\rho$.
The (blue) dots, (black) stars, (brown) squares, (red) crosses and (purple) diamonds denote the values we get numerically for $\alpha=1,2,3,4,5$ and the solid lines are for the associated large mass approximations.
\label{allplotF}}
\end{figure}

By the above procedure we have evaluated  with high precision  the renormalized effective action in given non-Abelian radial backgrounds. The results are shown in Fig.\ref{allplotF}.
In Fig.\ref{allplotF} we plot (as a function of $m\rho$) our numerically accurate results for the modified effective action $\tilde{\Gamma}$, defined in (\ref{renscale1}), with the choices $\alpha=1,2,3,4,5$ for the shape parameter in the radial profile function $f(r)$ in (\ref{backf}). We also plot, with solid lines, the predictions for the same quantity based on the large mass expansion (considered up to the order $\frac{1}{(m\rho)^4}$). The large mass expansion is generally quite good when $m\rho$ is large. But the validity range of the large mass expansion varies with the value of $\alpha$; as $\alpha$ becomes large, the large mass expansion is reliable only when $m\rho$ is significantly larger than the corresponding value of $m\rho$ with small $\alpha$. This is understandable, for the `small' quantity in the large mass expansion is really the ratio of typical values of 
the fields and the derivatives of fields to the mass. The derivatives have larger values when $\alpha$ becomes large, and thus the ratio becomes small only when the mass has a comparatively larger value.

\begin{table}[ht]
\caption{\label{table1}Table of two parameters in (\ref{smallmass}) for the five different values of $\alpha=1,2,3,4,5$.}
\begin{tabular}{cccccc}
\hline \hline $\alpha$& 1 &2 &3&4&5\\
\hline  $A_1$ & 0.145873 &   0.129759 & 0.113632  & 0.082787  &0.037283   \\
$A_2$ & 0.499(17) & 0.305(81) &   0.291(68)   & 0.292(11) & 0.295(03) \\
\hline\hline
\end{tabular}
\end{table}
We now comment on the case with $m=0$. In this case our numerical approach is not directly applicable. Solutions to (\ref{nonlinear}), $S_{(l,l+1/2)}(r)$ and $S_{(l+1/2,l)}(r)$,
behave like $\ln r$ and $-\ln r$ in the $r\to\infty$ limit, so the sum is finite. Analytic solutions to these equations, for the case of $\alpha=1$, can be found in \cite{idet}. In this instanton case, the $m\to 0$ limit was shown to be smooth, and furthermore to reproduce exactly 't Hooft's analytic massless result. We have found here that for other values of $\alpha$, the $m\to 0$ limit is once again smooth, and from the numerical values of the effective action for $m=\frac{1}{100}, \frac{2}{100}, \cdots, \frac{10}{100}$, we have found the following numerical extrapolations. In the case of $\alpha=1$ the extrapolation is
\begin{eqnarray}\label{smallmass}
\tilde\Gamma(m)&=& A_1 + A_2\, m^2 \,\ln m  \nonumber \\
&=& 0.145873(29) +  0.499(17) m^2 \ln m
\end{eqnarray}
In this case, analytic expressions for two leading coefficients in the small mass expansions are known, with the results:
$A_1=-\frac{17}{72}+\frac{1}{6}(1-\ln2)-2\zeta'(-1)\approx 0.14587331\cdots  $ \cite{thooft}; and $A_2= \frac{1}{2}$ \cite{carlitz,kwon}.  Note the remarkable agreement between these analytic results and the numerical extrapolation in (\ref{smallmass}).  Numerically determined (through extrapolation) values of $A_1$ and $A_2$ for the cases with other values of $\alpha$ are presented in Table \ref{table1}.  
We have not yet found simple analytic expressions for these leading small mass terms, although we suspect this should be possible.

\section{Quasi Abelian background fields}\label{section4}
\begin{figure}
\psfrag{hor}{$r$}
\psfrag{rzero}{$R_0=0$}
\psfrag{rone}{$R_0=1$}
\psfrag{rttwo}{$R_0=2$}
\psfrag{rthre}{$R_0=3$}
\psfrag{rfive}{$R_0=5$}
\psfrag{rrten}{$R_0=10$}
\psfrag{ver}{$g(r)$}
\subfigure[Plots of $g(r)$ with fixed $\beta=1$]{\includegraphics[width=0.6\linewidth]{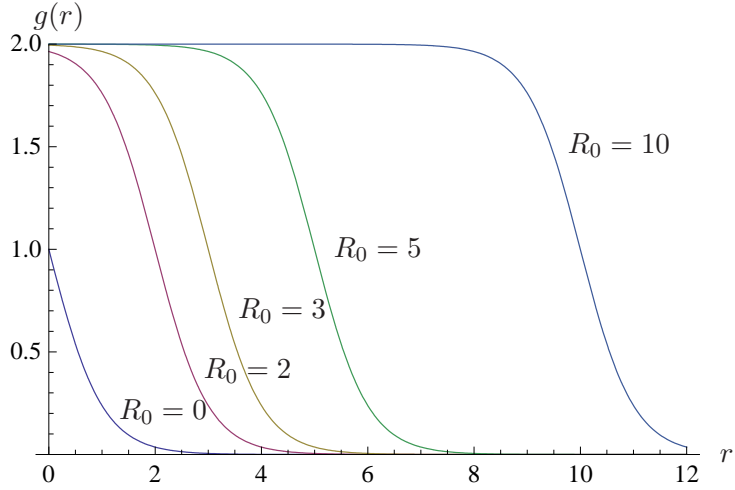}} 
\psfrag{hor}{$r$}
\psfrag{ver}{$g(r)$}
\psfrag{bfiv}{$\beta=5$}
\psfrag{bone}{$\beta=1$}
\psfrag{btwo}{$\beta=2$}
\psfrag{half}{$\beta=1/2$}
\psfrag{bfif}{$\beta=1/5$}
\subfigure[Plots of $g(r)$ with fixed $R_0=3$]{\includegraphics[width=0.6\linewidth]{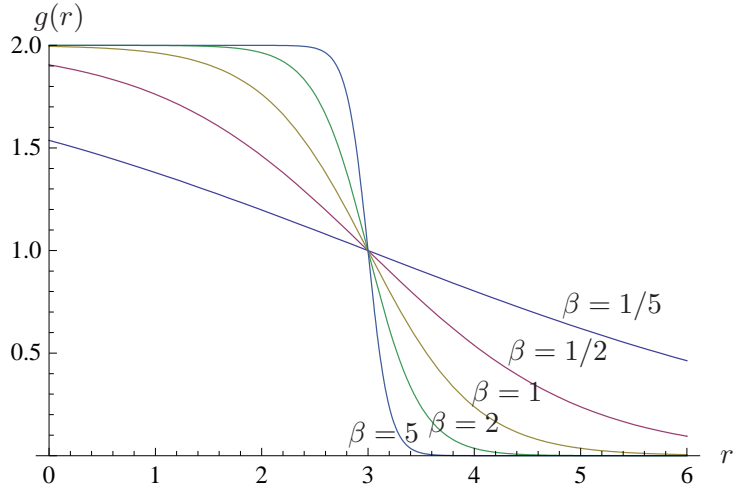} }
\caption{Plots of the function $g(r)$ (in units of $B$) for  various values of $R_0$ and $\beta$. In (a), plots are drawn for $R_0=1,2, 3, 5, 10$ with fixed $\beta=1$. In (b), plots are drawn for $\beta=5,2,1,1/2,1/5$ when $R_0=3$ \label{gplot}}
\end{figure}

In this section we consider the cases of quasi Abelian background fields, where an Abelian gauge field is embedded in the SU(2) Yang-Mills gauge fields. Using an appropriate isospin rotation we can take the non-vanishing components only in the 3rd direction. Then the vector potential can be written as  in (\ref{acase2}), and we take the radial profile function $g(r)$ in the step-like form (\ref{backg}).
Recall that there are three arbitrary parameters $\beta$, $\xi_0$, and $B$. When $\beta=0$ it corresponds to the case of uniform field strength studied in \cite{radial}.  When  $\beta\neq 0$, this potential describes a bubble shape  with a radius $R_0=\xi_0/\sqrt{B}$, and the dimensionless parameter $\beta$ is related to the thickness of the bubble. Illustrative graphs of the radial function $g(r)$ for various values of $R_0$ are drawn in Fig.\ref{gplot}.

The field strength tensor of the quasi Abelian gauge fields is
\begin{equation}\label{fmunucase2}
F_{\mu\nu}= -2 \eta_{\mu\nu i}\hat{u}^i g(r) \tau^3 + \frac{x_\lambda}{r}
(x_\mu \eta_{\nu\lambda i} - x_\nu \eta_{\mu\lambda i} ) \hat{u}^i g'(r)
\tau^3,
\end{equation}
The classical action can be evaluated as
\begin{eqnarray} \label{actioncase2}
\frac{1}{2}\int d^4x\, \text{tr}\, F_{\mu\nu}^2 &=&4 \pi^2
\int_0^\infty dr \; r^3 \left[ 8g(r)^2 + 4r g'(r)g(r) + r^2 g'(r)^2
\right] \\
&=&\frac{20 \pi^2}{\beta^4} \left({\rm PolyLog}(3, -e^{2 R_0}) -
   {\rm PolyLog}(5, -e^{2 R_0}) \right).
\end{eqnarray}
In the large $R_0$ limit, this quantity can be approximated as
\begin{eqnarray} \label{actionapprox}
\frac{1}{2}\int d^4x\, \text{tr}\, F_{\mu\nu}^2 \approx \frac{20\pi^2}{\beta^4}
\left(\frac{4}{3} R_0^5+\frac{10}{9}(\pi^2-6)R_0^3+\frac{1}{36}(7\pi^4-60\pi^2)R_0+\dots \right) \quad .
\end{eqnarray}

We turn to the evaluation of the one-loop effective action. When the background function is given in the form (\ref{backg}), based on dimensional considerations, we may cast the renormalized action into the form (\ref{renscale2}), and so simply compute $\tilde{\Gamma}\left(m/\sqrt{B}\right)$. Furthermore, since $\tilde{\Gamma}\left(m/\sqrt{B}\right)$ is a function just of $\frac{m}{\sqrt{B}}$, 
we may set $B=1$ hereafter, and it can be restored later.

\subsection{Large Mass and Derivative Expansions}

In the large mass expansion as given by (\ref{largemass2}), the modified effective action $ \tilde\Gamma(m)$ can be approximated by
\begin{equation}
\tilde\Gamma(m)_{\rm LM}=\tilde\Gamma^{(0)}_{\rm LM}\ln m + \tilde\Gamma^{(2)}_{\rm LM}\frac{1}{m^2}+
 \tilde\Gamma^{(4)}_{\rm LM}\frac{1}{m^4}+\cdots ,
\end{equation}
where
\begin{eqnarray}
\tilde\Gamma^{(0)}_{\rm LM}&=&-\frac{1}{12}\int_0^\infty dr
 r^3 \left(8 g(r)^2+4 r g(r) g'(r)+r^2 g'(r)^2\right)  \label{largemg1} \\
\tilde\Gamma^{(2)}_{\rm LM}&=&\frac{1}{720}\int_0^\infty dr
 \left[ 24 r^2 g(r) \left(15 g'(r)+r \left(9 g''(r)+r g^{(3)}(r)\right)\right) \right. \nonumber \\
&& \left. +r^3 \left(221 g'(r)^2+9 r^2 g''(r)^2+2 r g'(r) \left(71 g''(r)+6 r g^{(3)}(r)\right)\right)\right] \label{largemg2}\\
\tilde\Gamma^{(4)}_{\rm LM}&=& \frac{1}{10080}\int_0^\infty dr \left[
-2688 r^3 g(r)^4-540 g(r) g'(r)-2688 r^4 g(r)^3 g'(r)+595 r g'(r)^2 \right. \nonumber \\
&& -1456 r^5 g(r)^2 g'(r)^2-392 r^6 g(r) g'(r)^3-49 r^7 g'(r)^4+540 r g(r) g''(r)   \nonumber \\
&&+4480 r^2 g'(r) g''(r)+1837 r^3 g''(r)^2+1620 r^2 g(r) g^{(3)}(r) \nonumber \\
&& +2356 r^3 g'(r) g^{(3)}(r)+830 r^4 g''(r) g^{(3)}(r)+37 r^5 g^{(3)}(r)^2+504 r^3 g(r) g^{(4)}(r) \nonumber \\
&& \left. +380 r^4 g'(r) g^{(4)}(r)+52 r^5 g''(r) g^{(4)}(r)+18 r^4 \left(2 g(r)+r g'(r)\right) g^{(5)}(r)  \right]\label{largemg3}
\end{eqnarray}
For $g(r)$ given by the form (\ref{backg}) it is straightforward to evaluate numerically the integrals in (\ref{largemg1}), (\ref{largemg2})  and (\ref{largemg3}).

There is another well-known approximation method -- the derivative expansion . The leading term in this expansion is given in (\ref{dexpansion}),
where $\pm i E_1$ and $\pm i E_2$ are four  different eigenvalues of  the matrix $\mathbf{F}=(F_{\mu\nu})$, and 
in this quasi-Abelian case, can be expressed in terms of the function $g(r)$ as
\begin{equation}\label{eigenv}
E_1= 2 g(r)\quad , \quad E_2=2g(r)+r \frac{d\,g(r)}{d\,r}\quad .
\end{equation}
For the derivative expansion, one may thus insert the results (\ref{eigenv}) into our formula (\ref{dexpansion}) and {evaluate numerically the $r$ and $s$ integrals. This is also straightforward.}
These large mass and derivative expansion approximations will be compared with the exact results below.

\subsection{Numerically Accurate Calculation of the Lower Angular Momentum Part}

\begin{figure}
\psfrag{hor}{$r$}
\psfrag{ver}{$S_{l,l_3}(r)$}
\includegraphics[width=0.7\linewidth]{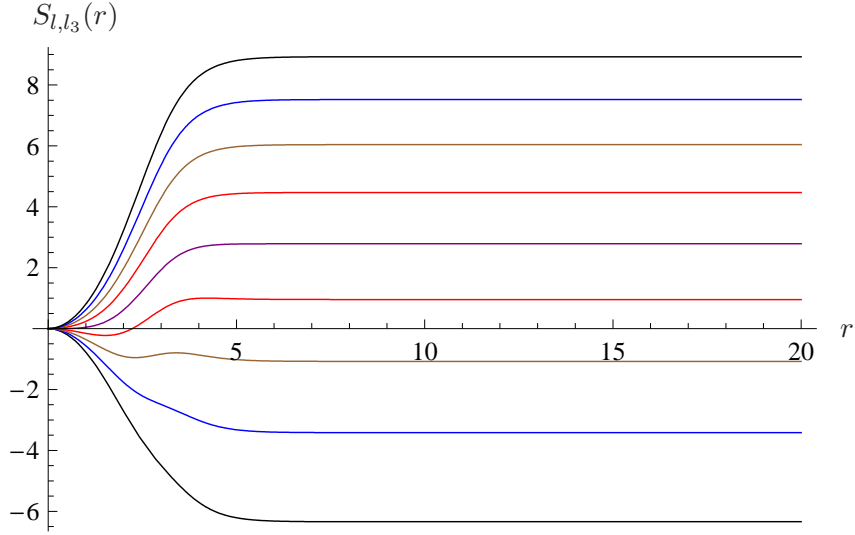}
\caption{Plot of solutions to (\ref{nonlinearcase2}) with $l_3=4,3,\cdots,-3,-4$ (from the top) when $l=4$, $\beta=1$,  $R_0=3$, and $m=1$. Note that the curves are essentially flat beyond $r\geq R_0$.}\label{solG}
\end{figure}

When the gauge vector fields have the form (\ref{acase2}), the partial waves are specified by the quantum numbers $J=(l,l_3,t_3,\bar{l}_3)$. The radial Hamiltonian becomes \cite{radial}
\begin{equation} \label{hcase2}
{\cal H}_{l,l_3,t_3} \equiv -D_{(l,l_3 , t_3 )}^2 = -\partial_{(l)}^2 +
8g(r) l_3 t_3 + r^2 g(r)^2
\end{equation}
Note that this Hamiltonian does not change  when we simultaneously change  the  sign of $t_3$ and $l_3$. Using this symmetry we can set $t_3=1/2$ without loss of generality.
Then the first (lower angular momentum) part of the renormalized action can be written as
\begin{equation}\label{lowcase2}
\Gamma_{J<J_L} (A;m) =2 \sum_{l=0}^{L}(2l+1)\sum_{l_3=-l}^{l}  \ln \left( \frac{ \det({\cal H}_{l,l_3,1/2}
+m^2)}{ \det({\cal H}_l^{\rm free} +m^2}) \right) .
\end{equation}
As in the case of Sec.\ref{section3} the ratio of two determinants is determined by the asymptotic value of
the function
\begin{equation}
S_{(l,l_3)}(r)=\ln \frac{\psi_{l,l_3}(r) }{\psi_l^{\rm free}(r)}
\end{equation}
which satisfies the following differential equation
\begin{eqnarray}
\frac{d^2 S_{(l,l_3)}}{dr^2}+\left(\frac{d S_{(l,l_3)}}{dr}\right)^2+\left(\frac{1}{r}+2m\frac{I^\prime_{2l+1}(m r)}{I_{2l+1}(m r)}\right)\frac{d S_{(l,l_3)}}{dr}=U_{(l,l_3)}(r)\quad ,
\label{nonlinearcase2}
\end{eqnarray}
with boundary conditions
\begin{eqnarray}
S_{(l,l_3)}(0)= S^\prime_{(l,l_3)}(0)=0 .
\label{logbc2}
\end{eqnarray}
The potential term $U_{(l,l_3)}(r)$ in (\ref{nonlinearcase2}) is given by
\begin{eqnarray}
U_{(l,l_3)}(r)= 4 l_3 g(r) + r^2 (g(r))^2.
\label{potential2}
\end{eqnarray}
This differential equation (\ref{nonlinearcase2}) can be solved numerically.
For the case with $\beta=1$ and $R_0=3$, we plot the solutions of (\ref{nonlinearcase2}) with $l_3=-4,-3,\cdots,3,4$ when $l=4$ in Fig.\ref{solG}.
For a given $l$, after summing all the values with $l_3=-l, \cdots, l$, we get the partial wave contributions:
\begin{equation}
P(l)= \sum_{l_3=-l}^{l} S_{(l,l_3)}(\infty).
\end{equation}
For large $l$ we find $P(l)$ behaving  like $\frac{1}{l}$.  

\subsection{WKB Calculation of the Higher Angular Momentum Part}

To compute the renormalized effective action we must identify the higher angular momentum part. The leading contributions in the large angular momentum limit is written in the form (\ref{largeL}). We present some of them in explicit forms:
\begin{eqnarray}
&& Q_{2}(r) = -\frac{8r^3g^2}{3\sqrt{\tilde{r}^2+4}}, \\
&& Q_{1}(r) = -\frac{2r^3(3\tilde{r}^2+8)}{(\tilde{r}^2+4)^{3/2}} g^2, \\
&& Q_{\rm log}(r) = -\frac{r^3}{12}\left( 8g^2+4rgg'+r^2g'^2 \right), \\
&& Q_{0}(r)= \frac{r^3}{90(\tilde{r}^2+4)^{7/2}} \left[
12r^4(5\tilde{r}^4+28\tilde{r}^2+32) g^4 -
30(9\tilde{r}^6+47\tilde{r}^4+40\tilde{r}^2+64) g^2 \nonumber \right. \\
&& \qquad \left. + 20r(3\tilde{r}^6+32\tilde{r}^4+88\tilde{r}^2+32)gg' +
5r^2(\tilde{r}^2+4)^2 \left\{ (3\tilde{r}^2+8)g'^2 - 8gg'' \right\} \right]
\nonumber \\
&& \qquad -Q_{{\rm log}}(r) \ln \left(\frac{\mu r}{\sqrt{\tilde{r}^2+4}+2}
\right), \\
&& Q_{-1}(r) = \frac{r^3}{4(\tilde{r}^2+4)^{9/2}}
\left[ 6r^4\tilde{r}^4(\tilde{r}^2+4) g^4 -(4\tilde{r}^8+7\tilde{r}^6+48\tilde{r}^4+1152\tilde{r}^2+1024) g^2 \right. \nonumber \\
&& \qquad \left. -16r(\tilde{r}^6+15\tilde{r}^4+52\tilde{r}^2+32) gg' - 4r^2(\tilde{r}^2+4)^2 \left\{ (\tilde{r}^2+2)g'^2 + \tilde{r}^2 gg'' \right\} \right],
\end{eqnarray}
with $\tilde{r}= \frac{m r}{L}$.
Higher order terms will be presented in \cite{hur}.

\subsection{Results for the Total One-loop Effective Action}

\begin{figure}
\psfrag{hor}{$m$}
\psfrag{title}{$R_0=0$}
\psfrag{ver}{$\tilde\Gamma(m)$}
\subfigure[]{\includegraphics[width=0.45\linewidth]{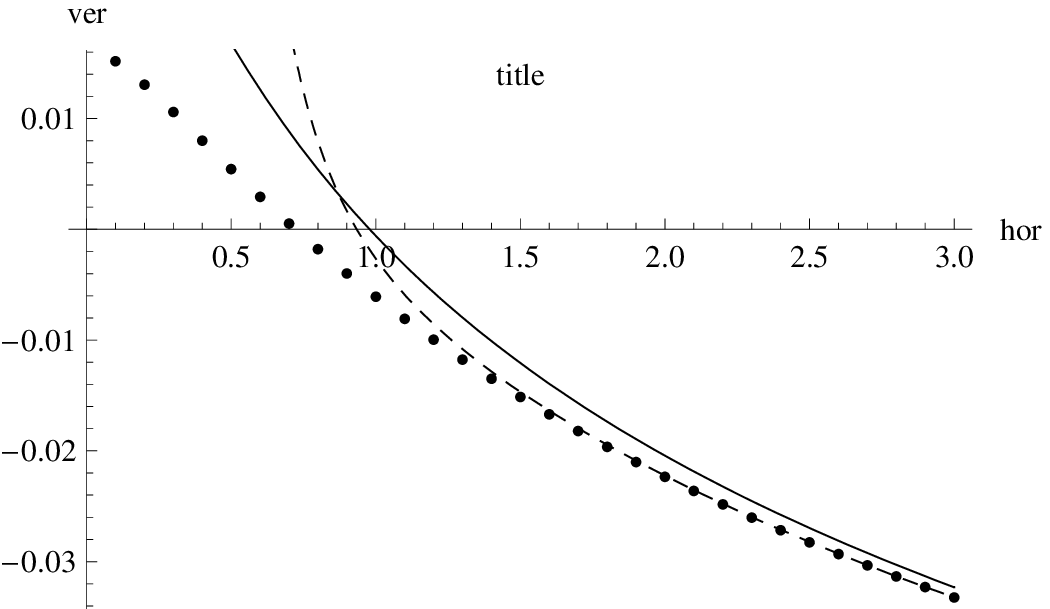}} 
\psfrag{hor}{$m$}
\psfrag{title}{$R_0=3$}
\psfrag{ver}{$\tilde\Gamma(m)$}
\subfigure[]{\includegraphics[width=0.45\linewidth]{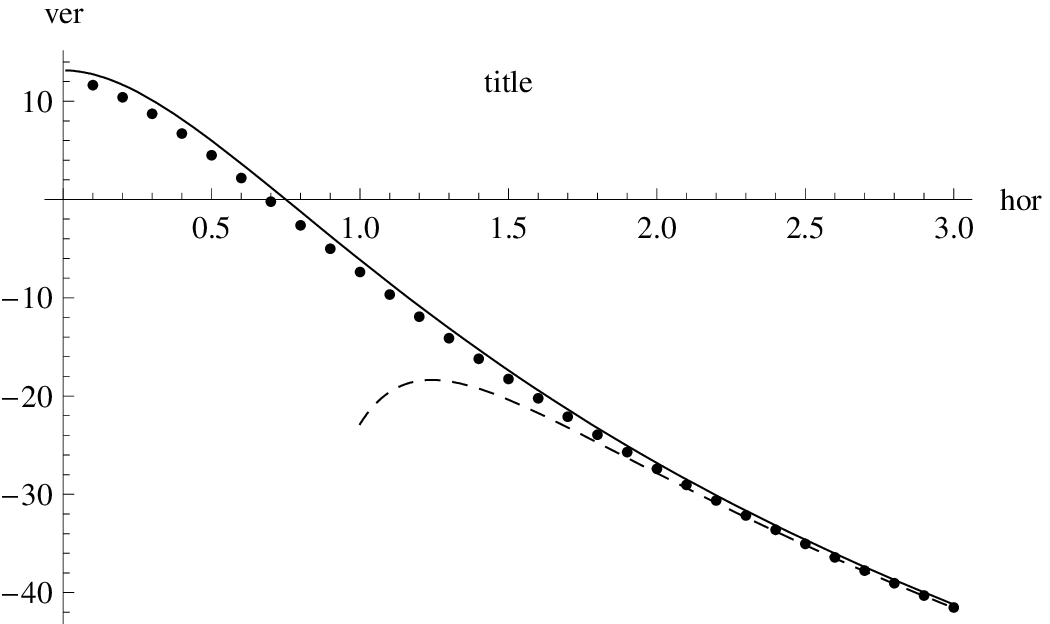}} \\
\psfrag{hor}{$m$}
\psfrag{title}{$R_0=5$}
\psfrag{ver}{$\tilde\Gamma(m)$}
\subfigure[]{\includegraphics[width=0.45\linewidth]{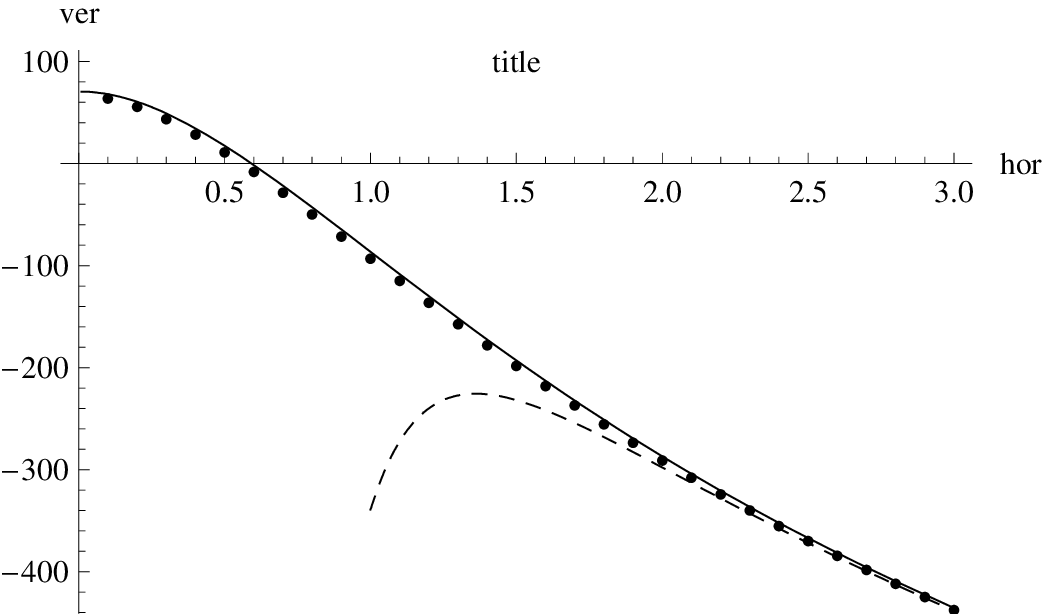}} 
\psfrag{hor}{$m$}
\psfrag{title}{$R_0=10$}
\psfrag{ver}{$\tilde\Gamma(m)$}
\subfigure[]{\includegraphics[width=0.45\linewidth]{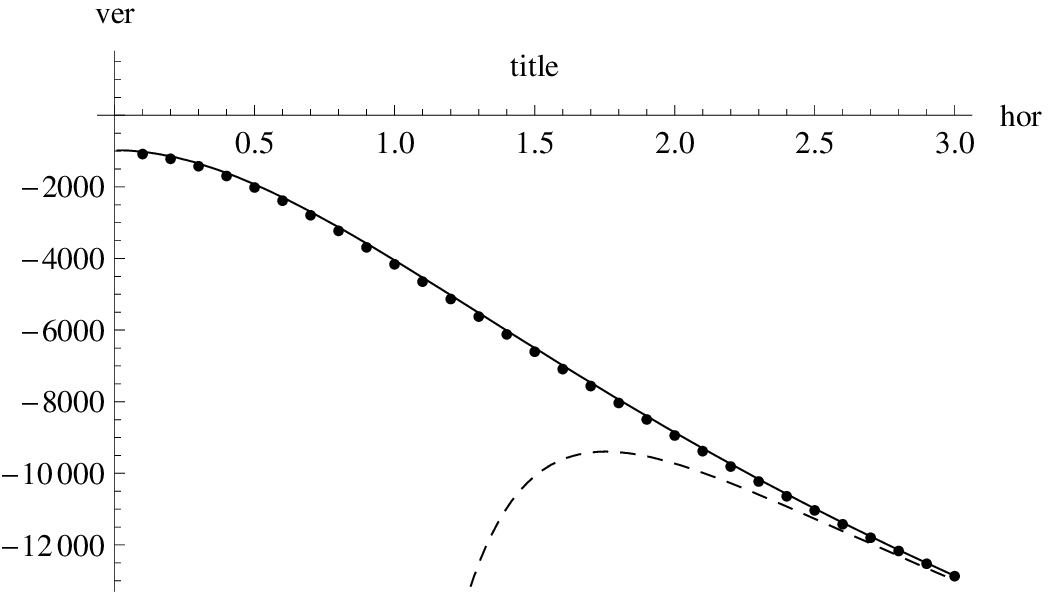}} 
\caption{Plots of the effective action for different $R_0$ values, i.e., $R_0=0$ in (a), $R_0=3$ in (b), $R_0=5$ in (c), and $R_0=10$ in (d), assuming $\beta=1$. The solid line in each figure denotes the result of the derivative expansion of the effective action while the dashed line denotes the result based on the large mass expansion. 
\label{actionG}}
\end{figure}
\begin{figure}[ht]
\psfrag{ver}{$\beta^4 \tilde\Gamma(m)$}
\psfrag{hor}{$m$}
\includegraphics[width=0.9\linewidth]{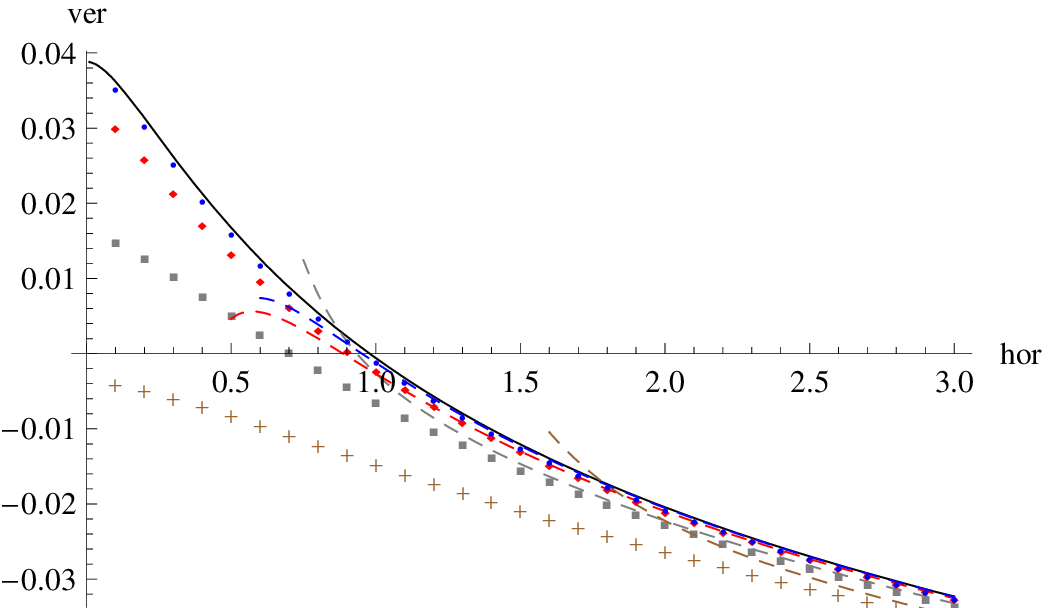}
\caption{Plot of the effective action multiplied by $\beta^4$ as a function of $m$ when $R_0=0$. The (blue) dots, (red) diamonds, (grey) squares and (brown) crosses denote the numerical values for $\beta=\frac{1}{5}$, $\beta=\frac{1}{2}$, $\beta=1$ and $\beta=2$, respectively. Each of dashed lines denotes the associated large mass expansion. The solid line denotes the leading derivative expansion which is independent of $\beta$ when $R_0=0$.
\label{plotG0beta}}
\end{figure}
As in the non-Abelian case treated in Sec.\ref{section3}, the large $L$ divergence of the numerical results for the low partial wave contribution is canceled by the large $L$ divergence in (\ref{largeL}), found analytically from radial WKB.
The combination of these lower and higher angular momentum parts is then  $L$-independent in the large $L$ limit,  as we have seen in the last section. In Fig.~\ref{actionG}, we plot the effective action for various values of the parameters $m, R_0$ with fixed value of $\beta=1$. 

In Fig.~\ref{plotG0beta}, we plot the effective action for various values of the parameters $m, \beta$ with fixed value of $R_0=0$. Its large mass expansion and its derivative expansion are drawn together for comparison. Clearly, the derivative expansion becomes more accurate when the parameter $\beta$, which represents of the derivative scale of the background function $g(r)$, becomes smaller. Note that the result of leading 
derivative expansion 
with $R_0=0$ becomes independent of $\beta$ when it is multiplied by $\beta^4$.  
Further note that the derivative expansion is generally much better than the large mass expansion for smaller values of $m$. This is despite the fact that we have just used the very leading order of the derivative expansion (\ref{dexpansion}), in which we simply take the Euler-Heisenberg constant field result, and then replaced the constant fields by their inhomogeneous forms in the effective Lagrangian. The superiority of the derivative expansion is because the derivative expansion is in fact a resummed version of the large mass expansion -- the large mass expansion is an expansion both in powers of the field and in derivatives of the field, while the derivative expansion is just an expansion in derivatives of the field, with {\it all} terms in the large mass expansion not involving derivatives having been resummed. This difference is clearly reflected in the plots shown in Fig.~\ref{actionG}.

\section{Conclusions}\label{section5}

To conclude, we have presented explicit computations of the renormalized one loop effective action for gauge field backgrounds that possess a radial symmetry, {such that the associated spectral problem can be decomposed into partial waves.} We considered one class of non-Abelian backgrounds, and another class of quasi-Abelian backgrounds, each characterized by a radial profile function. The computation has been performed using the partial wave cutoff method developed in \cite{idet,radial}, and we have further refined the numerical efficiency and precision by including higher order terms in the analytic radial WKB expression for the large partial wave contribution. 
The {main} conclusion is that the method works very efficiently and simply. With the incorporation of these higher order analytic terms, the numerical part of the computation is simplified because we do not need to take such high partial waves in the numerical computations. The method is not much more complicated to implement than the derivative expansion or the large mass expansion, and is much more accurate, especially in probing the small mass region. So, we can now reliably compute the renormalized determinant of any fluctuation problem whenever there is a radial symmetry. This method has now been tested successfully  in gauge theories \cite{idet,radial} and in self-interacting scalar field theories \cite{wipf,baacke,dunnemin,dunnewang}. The physical renormalization conditions are quite different in these various theories, but the WKB analysis {correctly} encodes the renormalization physics in each case. The method  has also motivated a recent extension of the Gel'fand-Yaglom theorem (for the determinant of ordinary differential operators), to partial differential operators that are radially separable \cite{dunnekirsten}.

An important generalization is to fermion fields. This can be done by converting the fermion problem to second order form, and then using the scalar method we have described here. {Aspects of this idea have been addressed in various approximation schemes \cite{carson,baacke2,strumia,burnier}; however, it should be possible to develop a more direct and numerically exact fermionic approach, along the lines of the scalar partial wave cutoff method described here.} 
Finally, while the method is 
restricted to backgrounds for which the fluctuation spectral problem is separable into partial waves, this includes a relatively large class of physically interesting cases, such as vortices, monopoles, and  instantons.

\section*{Acknowledgments} {GD thanks the DOE for support through grant DE-FG02-92ER40716.}
The work of CL was supported by the Korea Science Foundation ABRL
program (R14-2003-012-01002-0).

\end{document}